\documentstyle[aps,prl,multicol,psfig]{revtex}
\tighten
\begin{document}
\twocolumn[\hsize\textwidth\columnwidth\hsize\csname
@twocolumnfalse\endcsname
\draft
\title{Time-dependent Gutzwiller approximation for the Hubbard model}
\author{G. Seibold$^{\dagger}$ and J. Lorenzana$^{+,*}$ }
\address{$^{\dagger}$ Institut f\"ur Physik, BTU Cottbus, PBox 101344, 
         03013 Cottbus, Germany}
\address{$^+$ Istituto Nazionale di Fisica della Materia e
Dipartimento di Fisica, Universit\`a di Roma ``La Sapienza'',\\
Piazzale A. Moro 2, 00185 Roma, Italy}
\date{\today}
\maketitle
\begin{abstract}
We develop a time-dependent Gutzwiller  approximation (GA) for the Hubbard
model analogous to the time-dependent Hartree-Fock (HF) method.
The formalism incorporates ground state correlations of the random phase 
approximation (RPA) type beyond the GA.
Static quantities like ground state energy and double occupancy 
are in excellent agreement  with exact results in one dimension up to moderate 
coupling and in two dimensions for all couplings.  We find a substantial 
improvement over traditional GA and HF+RPA treatments. 
Dynamical correlation functions can be easily computed and are 
also substantially better than HF+RPA ones and obey well behaved sum rules.
\end{abstract}
\vskip2pc]
\narrowtext
The Gutzwiller (GW) trial wave function \cite{gut65} is probably the 
most popular variational approach to the Hubbard model which incorporates 
correlation effects beyond the Hartree-Fock (HF) approximation.
Since the Hubbard model describes the competition between hopping
and correlation induced localization of the charge carriers, 
the idea is to apply a projector to a Slater determinant (SD)
which reduces the number of doubly occupied sites. The optimum 
double occupancy probability is determined variationally.
 
Similar to  HF best results are obtained if one
allows for unrestricted charge and spin distributions which are determined
also variationally. For example for the half-filled Hubbard model a SD with 
  long range antiferromagnetic order is favoured\cite{yok87}.

A formal diagrammatic solution of the  GW variational problem has been given
by Metzner and Vollhardt\cite{met87,met88}, however,
for the most part of practical purposes one approximates the corresponding
expectation values using the so-called Gutzwiller approximation 
(GA)\cite{gut65}.

The GA can be derived using a variety of 
methods\cite{gut65,met87,met88,goe98,kot86}.
In particular  it is recovered at the mean-field level (saddle-point) of the
four-slave boson functional integral method introduced by Kotliar and
 Ruckenstein (KR)\cite{kot86}. 
The latter offers the possibility of going beyond
the Gutzwiller result as for example the 
inclusion of transversal spin degrees of freedom
\cite{li89}. In addition it provides a 
scheme to include fluctuations beyond the mean-field (MF) solution.
Expansions around the slave-boson saddle point have been performed
for homogeneous systems in Ref.~\cite{ras88,lav90} in order to
calculate correlation functions in the charge and longitudinal
spin channels.
However, the expansion of the KR hopping factor
$z^{SB}$ is a highly nontrivial task both with respect to
the proper normal ordering of the bosons and also with respect 
to the correct continuum limit of the functional integral \cite{arr93,arr95}.

The complexity of the expansions around the slave boson saddle point
have severely hampered practical computations of dynamical
quantities  within this formalism. One of the few successful attempts
is the computation of  the optical conductivity in the paramagnetic 
state in  Ref.~\cite{rai93,rai95}. Remarkably although the starting
SD describes a paramagnetic system, spectral  weight on the 
Hubbard bands appears as an effect of fluctuations. As far 
as we know this approach has not been pursued in broken symmetry states
due to  technical difficulties, including the fact that   
the KR choice for the $z^{SB}$ hopping factor does not lead to 
controlled sum rules\cite{raipc}. 

In this work we introduce a simple scheme to  compute fluctuation
corrections  around the GA to dynamical and static correlation functions
and the  ground state energy. The method can be viewed as a time-dependent
GA in the same way as the random phase approximation (RPA) method on top of
a HF solution (HF+RPA) can be viewed as time-dependent HF approximation in the 
limit of small amplitude oscillations \cite{rin80,bla86}. For this reason
we label the method as GA+RPA. 
It is also a generalization of the method of Ref.~\cite{vol84}
in order to describe the low temperature Fermi liquid regime.
The method incorporates  ground state correlations beyond
the ones of the Gutzwiller type just as HF+RPA takes into 
account ground state correlations not present in the HF wave function.

\begin{figure}
{\hspace*{0.cm}{\psfig{figure=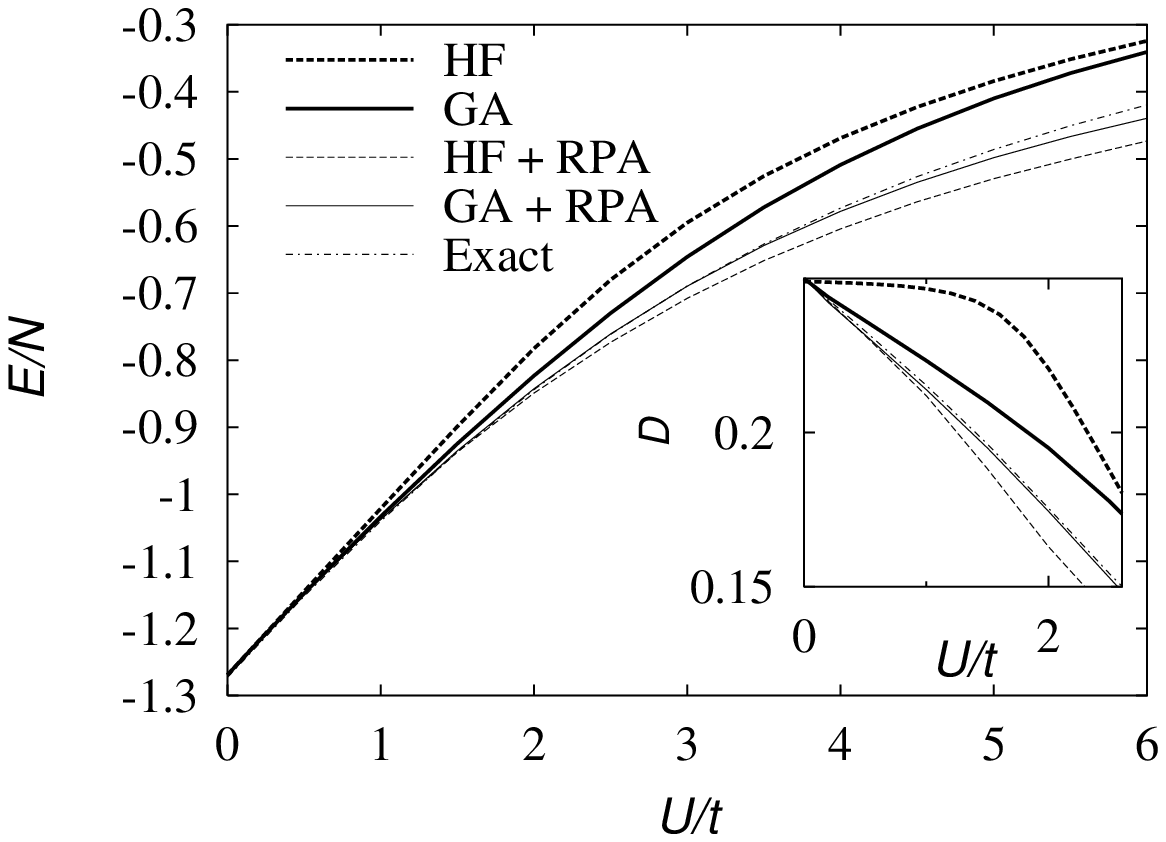,width=7cm}}}
{\hspace*{0.4cm}{\psfig{figure=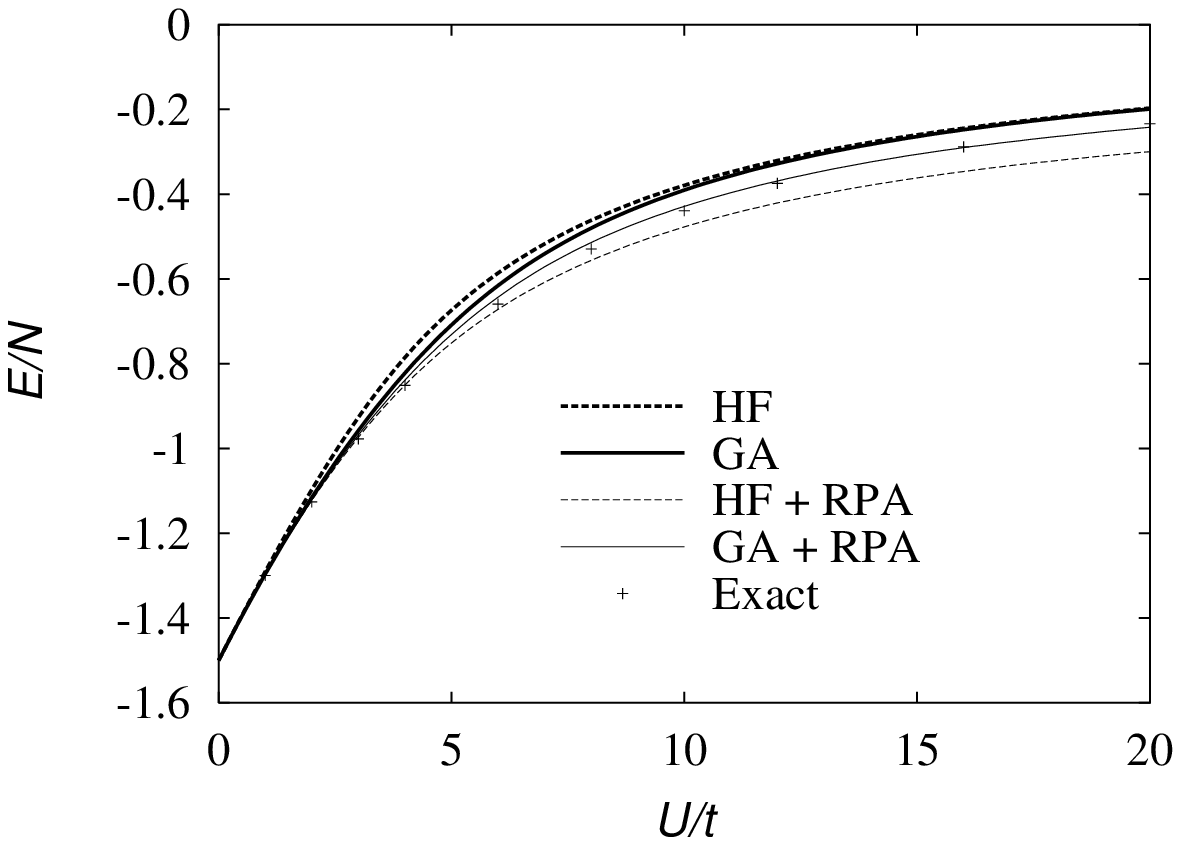,width=7cm}}}
\caption{ Comparison of the exact ground state energy  with the various  
approximate methods 
discussed in the text for the half-filled Hubbard model in 1d (upper panel)
and 2d (lower panel). 
 Exact results in the upper panel are for an infinite 1d system after
Ref.~\protect\cite{lie68} whereas approximate results are for a 32-site 
lattice (Finite size errors are estimated to be of the order of line width). 
Exact\protect\cite{fan90} and approximated results in the lower panel 
are for a 4x4 cluster.
The insets show the corresponding curves for the 1d double occupancy
(exact and GA+RPA are almost undistinguishable). 
}
\label{fig:edu}
\end{figure}

\begin{figure}
{\hspace{0cm}{\psfig{figure=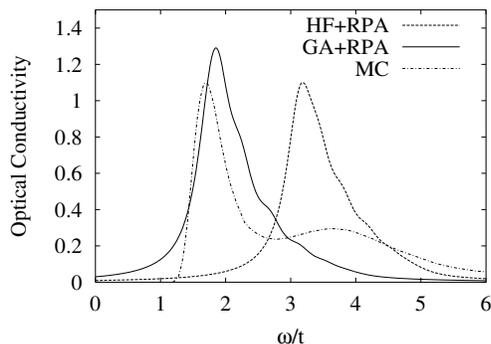,width=7.0cm}}}
\caption{ Optical conductivity of a 32-site Hubbard ring for $U/t=4$
in HF+RPA,  GA+RPA and  Monte-Carlo (MC) after Ref.~\protect\cite{per97}.
The individual peaks of the GA+RPA and HF+RPA data have been broadened
by a value of $0.25t$.} 
\label{fig:sdw}
\end{figure}

The GA+RPA ground state energy of the one-band Hubbard 
model is in  excellent agreement 
with exact results up to moderate coupling in one dimension (1d) and for 
all couplings in a 2d system (Fig.~\ref{fig:edu}). The optical conductivity 
of a Hubbard chain is in much better agreement with numerical results than 
HF+RPA (Fig.~\ref{fig:sdw}).  In addition sum rules are well behaved 
in the HF+RPA sense\cite{rin80,bla86,lor93a}. 

We consider the one-band Hubbard model 
\begin{equation}\label{HM}
H=\sum_{ij,\sigma}t_{ij} c_{i,\sigma}^{\dagger}c_{j,\sigma} + U\sum_{i}
n_{i,\uparrow}n_{i,\downarrow}
\end{equation}
where $c_{i,\sigma}$ destroys an electron 
with spin $\sigma$ at site
$i$, and $n_{i,\sigma}=c_{i,\sigma}^{\dagger}c_{i,\sigma}$. $U$ is the
on-site Hubbard repulsion and $t_{ij}$ denotes the transfer parameter between
sites $i$ and $j$. In the numerical computations below we take only
nearest neighbour matrix elements $t_{ij}=-t$ to be non-zero.

Our starting point is an energy functional $E[\rho,D]$ of the GA 
type\cite{goe98}. Here
$\rho$  is the  density matrix of an associated Slater determinant
$|SD>$, i.e. 
$\rho_{i\sigma,j\sigma'}=
\langle SD| c_{i\sigma}^{\dagger} c_{j\sigma'} |SD\rangle$ 
and $D$ is a vector of the  GA double occupancy parameters $D_i$  
at site $i$.
In order to consider arbitrary fluctuations the charge and spin 
distribution of $\rho$ and the distribution of $D$ should be completely
unrestricted.  For simplicity we consider only solutions 
where the associated  SD  is an eigenstate of the 
z-component of the total spin operator 
($\rho_{i\sigma,j\sigma'}\equiv \delta_{\sigma,\sigma'}\rho_{ij\sigma} $).

$E[\rho,D]$ can be obtained by exploiting the equivalence 
between the KR saddle point solution and the GA\cite{goe98}.
It is given by:
\begin{equation}\label{EGW}
E[\rho,D]=\sum_{ij\sigma}t_{ij}
z_{i\sigma}z_{j\sigma} \rho_{ij\sigma}
+U\sum_{i}D_{i} \label{E1}
\end{equation}
and
\begin{equation}
z_{i\sigma}=\frac{\sqrt{(1-\rho_{ii}+D_i)(\rho_{ii\sigma}-D_i)}
+\sqrt{D_i(\rho_{ii,-\sigma}-D_i)}}{\sqrt{
\rho_{ii\sigma}(1-\rho_{ii\sigma})}}.
\end{equation}
with  $\rho_{ii}=\sum_{\sigma}\rho_{ii\sigma}$.
The stationary solution ${\rho^{(0)},D^{(0)}}$ is determined by 
minimizing the energy functional with respect to $\rho$ and $D$.

The variation with respect to the density
matrix has to be constrained to the subspace of Slater determinants by
imposing the projector condition $\rho^2=\rho$\cite{rin80,bla86}.
Within this subspace we now consider small time-dependent amplitude 
fluctuations of the density matrix $\rho(t)$.   We add a weak 
time-dependent field to Eq.~(\ref{E1}) of the form:
$F(t)=\sum_{i\sigma,j\sigma'} (
f_{i\sigma,j\sigma'} e^{-i\omega t}  c_{i\sigma}^{\dagger} c_{j\sigma'} 
+h.c. )$.
This produces  small amplitudes oscillations $\delta\rho(t)$   
around the stationary density i.e. $\delta\rho(t)\equiv \rho(t)-\rho^{(0)}$. 

We  assume that at each instant of time the double occupancy 
parameter is at the minimum of the energy functional compatible
with the corresponding $\rho(t)$; i.e. the double occupancy
parameters $\{D\}$ adjust antiadiabatically to the time evolution 
of the density matrix. 
This is reasonable since the double occupancy involves 
processes which are generally high in energy and hence fast. 
We anticipate that for the cases explored,  this  approximation works 
well up to energies
as large as the  Hubbard band in the optical conductivity 
(Fig.~\ref{fig:sdw}). 
  As the density varies the double occupancy shifts from 
$D^{(0)}$  to satisfy the antiadiabaticity constraint. We define
  $ \delta D(t)=   D(t)-D^{(0)}$ 
and $\delta\rho(t)$ and $\delta D(t)$
are linear in $f$.

The formal complication of the present approach as compared to the
standard RPA has its origin in the proper adjustment of $D$ to
the time evolution of $\rho(t)$, i.e. the determination of  $\delta D(t)$.
This step is achieved by expanding the energy functional Eq.~(\ref{E1}) 
up to second order in $\delta \rho$ and $\delta D$ around the
saddle point:% (note that the linear variations vanish in this case):
\begin{eqnarray}
  \label{eq:e2}
E[\rho,D]&=&E_0+ \overline h_0^{\dagger}\delta\overline{\rho} 
+ \frac12 \delta\overline{\rho}^{\dagger} L_0  \delta\overline{\rho}\\
&+&
\delta  D  S_0  \delta {\overline\rho} 
+ \frac12 \delta D^t  K_0  \delta D\nonumber
 \end{eqnarray}
where the bar indicates that we are treating a matrix as a 
 column vector and the not indicates evaluation in the stationary
state. 
Here  we have defined  an effective one-particle Gutzwiller 
Hamiltonian\cite{rin80,bla86}:
\begin{equation}
  \label{eq:hgw}
  h_{ji\sigma}=\frac{\partial E } {\partial \rho_{ij\sigma}}
\end{equation}
and the matrices 
\begin{eqnarray}
  \label{eq:mat}
L_{ij\sigma,kl\sigma'}&=&\frac{\partial^2 E}
{\partial\rho^*_{ij\sigma}\partial\rho_{kl\sigma'}}\\
S_{k,ij\sigma}&=&\frac{\partial^2 E}
{\partial D_{k}\partial\rho_{kl\sigma }} \\
K_{k,l}&=&\frac{\partial^2 E}
{\partial D_{k}\partial D_{l}}.
\end{eqnarray}
 Using the condition of antiadiabaticity 
\begin{equation}\label{AAC}
\frac{\partial E}{\partial \delta D}=0
\end{equation}
in Eq.~(\ref{eq:e2})
we obtain a linear relation between $\delta \rho$ and $\delta D$.
Eliminating $\delta D$ from Eq.~(\ref{eq:e2}) finally yields an expansion 
of the 
energy as a functional of $\delta \rho$ alone   
$\widetilde E[\rho]\equiv E[\rho,D(\rho)]$,
\begin{eqnarray}
  \label{eq:e2rho}
\widetilde E[\rho]=E_0+ \overline h_0^{\dagger}\delta\overline{\rho} 
+ \frac12 \delta\overline{\rho}^{\dagger}  
(L_0 - S_0 ^{\dagger}  K_0^{-1} S_0) \delta\overline{\rho}
\end{eqnarray}
This can be regarded as the expansion of an effective interacting energy 
functional in which the interaction potential between particles  is  
density dependent. This kind of functional often appears in 
the context of nuclear physics and a well developped
machinery exist to compute the RPA fluctuations induced by the interaction.
We will only briefly outline here the corresponding formalism 
(for details see Ref.~\cite{rin80,bla86}). 
The advantage of this method with 
respect to other methods (eg. diagrammatic) 
is that the present derivation is solely based on the knowledge of 
an energy functional of a  SD density matrix
which is precisely what the GA provides.

At the saddle-point  $h$ and   $\rho$  can be 
diagonalized simultaneously. 
As a result one obtains $(h_0)_{kl}=\delta_{kl} \epsilon_k$
and the density matrix has eigenvalue 1 below the Fermi level 
and eigenvalue 0 above the Fermi level. We will notate
states below the Fermi level as hole ($h$) states and the states
above the Fermi level as particle states ($p$).

Up to linear order the density matrix obeys the equation of 
motion\cite{rin80,bla86}:
 \begin{equation}
   \label{eq:motion}
   i\hbar \delta \dot \rho=[\widetilde  h_0,\delta\rho]+
[\frac{\partial\widetilde h}{\partial \rho}.\delta \rho, \rho^{(0)}]
+[f^{GA},\rho^{(0)}]
 \end{equation}
where $\widetilde h$ is defined as in Eq.~(\ref{eq:hgw}) but with 
$\widetilde E$ instead of $E$ (Note that $\widetilde  h_0=h_0$).
$\frac{\partial\widetilde h}{\partial \rho}.\delta \rho$ is a short hand 
notation for
\begin{equation}
  \label{eq:notation}
  \sum_{ph}
\left(\left. \frac{\partial\widetilde h } 
           {\partial \rho_{ph}}\right|_{\rho=\rho^{(0)}}\delta \rho_{ph}+
      \left.\frac{\partial\widetilde h}{\partial \rho_{hp}}\right|_{\rho=\rho^{(0)}}
\delta \rho_{hp}\right).
\end{equation}
$f^{GA}$ is the GA version of $f$, i.e. it includes the 
$z_0$ factors for intersite matrix elements. 

One can show that particle-particle and hole-hole matrix elements of 
Eq.~(\ref{eq:motion}) are zero and the particle-hole ($ph$) matrix elements 
of $\delta \rho$ satisfy the well known RPA eigenvalue equation. The 
RPA dynamical matrix can be obtained from 
Eqs.~(\ref{eq:e2rho}),(\ref{eq:motion}).
 Upon diagonalizing the RPA matrix
 by a Bogoliubov transformation one obtains the eigenvectors
$V^{(\lambda)}=(X^{(\lambda)}_{ph}, Y^{(\lambda)}_{ph})$ and eigenvalues 
$E^{(\lambda)}$
where the latter correspond to the excitation energies of the system.
An explicit expression for the response functions and a discussion of sum 
rules can be found in Ref.~\cite{rin80,bla86} which apply 
straightforwardly to our case. 

The present formalism  is well suited  for the  calculation of charge 
excitations in 
inhomogeneous doped systems\cite{goe98,lor93a,yon93,goe981} which will be 
presented elsewhere. In the following we restrict ourselves to the 
half-filled Hubbard model in the antiferromagnetic N\'{e}el state\cite{yok87}. 
The  double occupancy at the RPA level is given by:
$D_{RPA}=\int d\omega \sum_{\lambda}
\langle 0| n_{i\uparrow}|\lambda \rangle 
\langle \lambda | n_{i\downarrow}|0 \rangle 
\delta(\omega-E^{(\lambda)}) $ where the integrand is the Lehmann 
representation of an appropriately defined density-density correlation 
function.  The matrix elements $\langle 0| n_{\uparrow}|\lambda \rangle $
for $\lambda>0$ can be computed in terms of the eigenvectors 
$V^{(\lambda)}$\cite{rin80,bla86}.

In the inset of  Fig.~\ref{fig:edu}, we show the GA+RPA double occupancy 
compared with exact results and other approximations in a 1d system.
For small $U$ long range magnetic order is not enough to reduce
substantially the HF double occupancy from the noninteracting value. 
RPA on top of HF corrects for this but because the starting point is 
quite far from the exact result the correction is not so accurate and 
one gets that HF+RPA  overstrikes the exact double occupancy. 
On the contrary for the GA only a small correction is 
needed and RPA performs remarkably well. Note that 
$U (D_{RPA} - D_{HF})$ is a  measure for the residual interaction in HF+RPA.
 In the GA+RPA approach such a simple relation is lost but clearly a smaller
 correction of the MF double occupancy suggests a smaller residual 
interaction.

From the interaction energy $U D_{RPA}$ we compute the correction 
to the ground state energy using the coupling constant integration 
trick\cite{mah90}.  We find very good agreement 
with the exact results as shown in Fig.~\ref{fig:edu}. This holds in 1d 
up to  intermediate values of $U/t$ and in a 4x4 2d cluster for all $U/t$. 
The improvement  with dimensionality is expected as in any 
MF + RPA computation. 
In order to examine the quality of dynamical correlation functions 
we have studied the optical conductivity in the GA+RPA approach. 
 As in the HF+RPA method~\cite{rin80,bla86,lor93a}
the f-sum rule is exactly satisfied with the following prescription.
The optical conductivity
on one side of the equality should be computed at the GA+RPA 
(HF+RPA) level and the expectation  value on the other side 
(essentially the kinetic energy in our case\cite{sha90}) 
computed at GA  (HF) level. 

In this regard the f-sum rule provides also an encouraging argument that the
GA+RPA dynamical correlation functions are much more accurate than those
obtained via the corresponding HF+RPA method.
We have compared  the exact kinetic energy  of a 4x4 lattice\cite{dag92}
with unrestricted GA and HF results for various 
hole concentrations and have found that over a wide range of doping and 
on-site correlation $U$ there is almost perfect agreement between the 
GA method and exact results. On the other hand the HF kinetic energy
has an error that for example for $U=4t$ is at least 40 times larger.
This is not surprising since GA takes into account the correlation 
induced reduction of kinetic energy in a much better way than HF. 

Fig.~\ref{fig:sdw} displays $\sigma(\omega)$ for a 32-site Hubbard ring and 
half-filling in case of $U/t=4$.
The onset of excitations across the Mott-Hubbard gap is signalled by
the appearance of a large peak in $\sigma(\omega)$. We find excellent
agreement between  Monte-Carlo (MC) and GA+RPA 
whereas the excitation energy in HF is clearly overestimated. 
Note that the MC data display
an additional hump at approximately twice the energy of the first peak.
Both particle-particle scattering processes (not included at RPA level)
and the failure of the antiadiabaticity assumption for the double occupancy 
at high energies can explain the discrepancy. We see however that for energies
 of the order of the   Hubbard band the method performs very well. 

In conclusion we have presented a time-dependent GA 
for the calculation of dynamical and static quantities in the Hubbard model.
The approach is conceptually very simple and leads to  
much better agreement with exact results than previous approximations. 

As in any computation of fluctuations we are dealing with the residual 
interaction between particles beyond the mean-field level.
Roughly speaking since the GA contains ground state 
correlations not included in a HF wave function  the  
residual interaction is a smaller perturbation to the MF state
and hence it is natural that RPA works much better in this case. 

It is interesting to remark that the computation of the ground state energy
presented here is reminiscent of the evaluation of the one
in the uniform electron gas based on Hubbard type 
dielectric functions\cite{mah90}. Also there the computation goes through a 
density-density correlation function and the coupling constant 
integration trick. Furthermore the Hubbard local field correction to the 
dielectric function takes into account the correlation hole in the uniform
electron gas whereas the GW projector method  takes into account
similar correlations in the Hubbard model. The connection 
between these two approaches  deserves further investigation as it may
lead to a unified approach to strongly correlated systems. 

We greatfully acknowledge many useful discussions with C. Di Castro, 
C. Castellani, M. Grilli and R. Raimondi.  We acknowledge
partial financial support from INFM. G.S. acknowledges financial support 
from the 
Deutsche Forschungsgemeinschaft as well as  hospitality and 
support from the Dipartimento
di Fisica of Universit\`a di Roma ``La Sapienza''where part of this work
was carried out.

\end{document}